\documentstyle[12pt]{article}

\newtheorem{theorem}{Theorem}

\def\qed{\hfill{\vrule height 3mm width 2mm depth 0mm}}

\def\be{\begin{equation}}
\def\ee{\end{equation}}
\def\bea{\begin{eqnarray}}
\def\eea{\end{eqnarray}}
\def\lf{ \nonumber\\ }

\def\IR{\hbox{{{I}\kern-0.25em{R}}}}

\def\phip{{\phi^\prime}}
\def\phib{{\phi^{\prime\prime}}}

\def\xmu{{x^\mu}}
\def\xnu{{x^\nu}}
\def\xsi{{x^\si}}

\def\ymu{{y^\mu}}
\def\ynu{{y^\nu}}
\def\zmu{{z^\mu}}

\def\si{\sigma}

\def\tr{{\hbox{{\rm tr}}}}

\def\d{\partial}

\def\dd#1{{\d \over {\d #1}}}
\def\ddd#1#2{{{\d #1} \over {\d #2}}}

\def\Jac#1#2{{\bv \ddd #1 #2 \bv}}

\def\bv{{\bigg\vert}}

\begin{document}
\pagestyle{empty}

\title{Diffeomorphism group and conformal fields}

\author{
T. A. Larsson\thanks{
Supported by the Swedish Natural Science Council (NFR). } \\
Dept. of Theoretical Physics \\
100 44 Stockholm \\
Sweden \\
Email address: tl@theophys.kth.se \\
\date {August 1992}
}

\maketitle

\begin{abstract}
Conformal fields are a new class of $Vect(N)$ modules which
are more general than tensor fields. The corresponding
diffeomorphism group action is constructed.
Conformal fields are thus invariantly defined.
\end{abstract}

\vglue 2 cm
PACS numbers: 02.20, 02.40
\vglue 1 cm
hep-th/9208043
\newpage
\pagestyle{plain}
\setcounter{page}{1}

{\em 1.}
Any meaningful object living in $N$-dimensional space must be
invariantly defined, i.e. it must transform as a representation
of $Diff(N)$, the diffeo\-morphism group in $N$ dimensions.
$Diff(N)$ representation theory therefore amounts to
a classification of inequivalent meaningful objects, which
appears to be of importance both in physics and geometry.
In particular, the main concepts of differential geometry (tensor
fields, connections, exterior derivatives) have a natural
description in this language. This point of view has mainly
been advocated by Russian mathematicians \cite{Rud74} --
\cite{Fuks87}.

We have recently \cite{Lar92a} \cite{Lar92b} discovered a
new class of modules of $Vect(N)$, the algebra of vector fields in
$N$ dimensions, which is the Lie algebra of $Diff(N)$. These
modules, which were named {\em conformal fields}, are in a
sense more natural than tensor fields.
However, $Vect(N)$ is not very interesting in its own right, but
only as the infinitesimal version of $Diff(N)$.
Therefore, it is important to check that conformal fields
can be exponented to $Diff(N)$ representations. This is the
subject of this letter.

It is known \cite{Rud74} -- \cite{Fuks87} that tensor fields
can be described as those $Diff(N)$
modules which are induced from the group $GL(N)$ consisting of
the linear transformations
\be
\xmu \mapsto a^\mu_\nu \xnu.
\ee
However, the largest finite-dimensional subgroup of $Diff(N)$
is the group of projective transformations,
\be
\xmu \mapsto {{ a^\mu_\nu \xnu + a^\mu_0} \over
{ a^0_\nu \xnu + a^0_0}}.
\label{proj}
\ee
This group is isomorphic to $SL(N+1)$ because
two consecutive projective transformations
correspond to multiplication of the matrices
\be
a^A_B \equiv
\pmatrix{
a^0_0 & a^0_\nu \cr
a^\mu_0 & a^\mu_\nu \cr
}.
\ee
Since this matrix is only determined by (\ref{proj}) up to an
over-all factor, we may put $\det a = 1$.

We thus have the following inclusions of groups
\be
GL(N) \subset SL(N+1) \subset Diff(N).
\ee
A conformal field\footnote
{A better name would be ``projective
field''. The prefix ``conformal'' was chosen because we did not
recognize the Lie algebra of the projective group, which is
somewhat similar to the conformal algebra.}
is a $Diff(N)$ module which is induced from the projective group.
Because of a theorem on induction in stages \cite{Kir76},
this definition includes tensor fields: induce from an
$SL(N+1)$ module which in turn has been induced $GL(N)$. However,
if we start from an $SL(N+1)$ module which is not induced from
$GL(N)$, a new kind of invariantly defined object arises.

The above description of conformal fields is not
very concrete. On the Lie algebra level, we can be much more
explicit.
$Vect(N)$ reads
\be
[L_f, L_g] = L_{[f,g]},
\ee
where $L_f$ is the Lie derivative corresponding to the vector
field $f = f^\mu \d_\mu$ ($\d_\mu \equiv \d / \d \xmu$).
Tensor fields are defined by
\be
L_f = f^\mu \d_\mu + \d_\mu f^\nu T^\mu_\nu
\label{tens}
\ee
where $T^\mu_\nu$ are the generators of $gl(N)$, i.e.
\be
[T^\mu_\nu, T^\si_\tau] = \delta^\si_\nu T^\mu_\tau
- \delta^\mu_\tau T^\si_\nu.
\label{gl}
\ee
By substituting different representations of $gl(N)$ into
(\ref{tens}), we obtain all kinds of tensor fields.

It should be noted that upper and lower indices have been
switched, compared to our previous papers. This change in
notation was performed in order to agree with other literature.

To describe conformal fields, it is useful to introduce the
following notation.
Let $A = (0, \mu)$ be an  $N+1$-component index, and let
\be
\begin{array}{rclcccrcl}
f^A &=& (0, f^\mu), & & & &
x^A &=& (t, \xmu), \\
\\
\d_B &=& (- t^{-1} \xsi \d_\si, \d_\nu), & & & &
k_B &=& (t^{-1}, 0),
\end{array}
\label{notation}
\ee
where $t$ is a parameter.
{}From (\ref{notation}) we deduce the following useful formulas.
\be
\begin{array}{rclcccrcl}
[\d_A, \d_B] &=& k_A \d_B - k_B \d_A, &&&&
[\d_B, x^A] &=& \delta^A_B - k_B x^A, \lf
\lf
x^A \d_A &=& k_A f^A = 0, &&&&
k_A x^A &=& 1.
\end{array}
\ee
Now, the following expression satisfies $Vect(N)$ \cite{Lar92a}.
\be
L_f = f^A \d_A +
\Big\{ (\d_A + k_A) f^B + c x^B \d_A \d_C f^C \Big\} T^A_B,
\label{conf}
\ee
where $c$ is another parameter and $T^A_B$ are the
generators of $sl(N+1)$, i.e.
\be
[T^A_B, T^C_D] = \delta^C_B T^A_D - \delta^A_D T^C_B,
\qquad \qquad
T^A_A = 0.
\ee

{\em 2.}
We now present the analogs of (\ref{tens}) and (\ref{conf})
at the group level. Under the diffeomorphism
$x \to y(x)$, a $Diff(N)$ representation $\phi(x)$ transforms
as $\phi(x) \to \phip(x)$. We say that this representation
is {\em field-like}, if
\be
\phip(y) = U(y,x) \phi(x),
\label{field}
\ee
where $\phi$ is a vector and $U$ is a matrix in some
finite-dimensional vector space.
Under two consequtive diffeomorphisms, $x \to y \to z$,
the field transforms as $\phi \to \phip \to \phib$,
where
\be
\phib(z) = U(z,y) \phip(y) = U(z,y) U(y,x) \phi(x).
\ee
Comparing with (\ref{field}), we see that the matrices
$U(y,x)$ must fulfil
\be
U(z,y) U(y,x) = U(z,x),
\qquad\qquad
U(x,x) = 1.
\label{rep}
\ee
Clearly, the dual field, which transforms as
\be
\phip(y) = \phi(x) U(x,y),
\label{dual}
\ee
is also a $Diff(N)$ representation.
Tensor products of (\ref{field}) and (\ref{dual}) can also
be considered.

The classical field-like objects are tensor fields,
\be
U^\mu_\nu(y,x) = \ddd \ymu \xnu,
\label{tensor}
\ee
because
\be
\ddd \zmu \ynu \ddd \ynu \xsi = \ddd \zmu \xsi,
\qquad
\ddd \xmu \xnu = \delta^\mu_\nu.
\ee
Eq. (\ref{rep}) also admits a one-dimensional representation,
which corresponds to densities with weight $\lambda$. Let
\be
\Jac y x \equiv \det \bigg( \ddd y x \bigg)
\label{jac}
\ee
be the Jacobian of the mapping $x \to y$. Then
\be
U(y,x) = {\Jac y x}^{-\lambda}
\ee
satisfies the representation conditions (\ref{rep}).

Our main result is the following field-like representation of
$Diff(N)$.

\begin{theorem}:
Let $c$ be a parameter, and let $x^A$, $\d / \d{x^B}$ and $k_B$
be defined as in (\ref{notation}).
Then the following $(N+1) \times (N+1)$-dimensional matrices
satisfy (\ref{rep}).
\be
U^A_B(y,x)
= \Big\{ \dd{x^B} + k_B + c \dd{x^B}h(y,x) \Big\}  y^A,
\label{matrix}
\ee
where
\be
h(y,x) = \log \Jac y x = \tr\ \log \bigg( \ddd y x \bigg).
\ee
\label{Thm1}
\end{theorem}

More explicitly, the matrix reads
\be
U^A_B(y,x) = \pmatrix{
U^0_0 & U^0_\nu \cr
U^\mu_0 & U^\mu_\nu
}
= \pmatrix{
1 - c \xsi \d_\si h
& c t \d_\nu h \cr
t^{-1} \big( \ymu - \xsi \d_\si \ymu - c \ymu \xsi \d_\si h \big)
& \d_\nu \ymu + c \ymu \d_\nu h
}.
\ee

Note that $| \d y / \d x |$ is the determinant of the
$N \times N$-dimensional matrix $\d \ymu / \d \xnu$;
because $y^0 \equiv t$, the determinant of $\d y^A / \d x^B$
vanishes.
The corresponding fields $\phi^A(x)$, $\phi_B(x)$, and tensor
products thereof, are called {\em conformal fields}. Of course,
we can also consider tensor products of conformal fields with
tensor fields or densitites.

{\em Proof of Theorem \ref{Thm1}:}
We begin by noting that
\be
\dd{x^B} = \ddd{y^C}{x^B} \dd{y^C}.
\label{cc}
\ee
Namely, the LHS equals
\be
(-{\xmu \over t}, 1) \dd\xmu
= (-{\xmu \over t}, 1) \ddd\ynu\xmu \dd\ynu,
\ee
whereas the RHS is
\be
(-{\ynu \over t}, 1) \pmatrix{
0 & 0 \cr
\cr
\displaystyle{-{\xmu \over t} \ddd \ynu \xmu}
& \displaystyle{\ddd \ynu \xmu} }
\dd \ynu.
\ee

The theorem asserts that
\be
U^A_B(z,x)
= \ddd{z^A}{x^B} + k_B z^A + c \dd{x^B} h(z,x) z^A
\ee
and
\bea
\lefteqn{ U^A_C(z,y) U^C_B(y,x) } \lf
&=& \Big( \dd{y^C} + k_C + c \dd{y^C} h(z,y) \Big) z^A \;
\Big( \dd{x^B} + k_B + c \dd{x^B} h(y,x) \Big) y^C \lf
&=& \ddd{y^C}{x^B} \ddd{z^A}{y^C} + k_B z^A
+ c \Big( \ddd{y^C}{x^B} \dd{y^C} h(z,y) z^A
+ z^A \dd{x^B} h(y,x) \Big)
\eea
are equal. Using
\be
k_C \ddd{y^C}{x^B} = y^C \dd{y^C} = 0,
\qquad
k_C y^C = 1,
\ee
and (\ref{cc}) we see that the terms independent of $c$ agree.
The terms linear in $c$ yield
\be
\dd{x^B} h(z,x) = \dd{x^B} h(z,y) + \dd{x^B} h(y,x),
\ee
i.e.
\be
h(z,x) = h(z,y) + h(y,x).
\ee
That this functional equation is solved by
\be
h(y,x) = \log \Jac y x
\ee
follows from
\be
\Jac z y \; \Jac y x = \Jac z y.
\ee
Finally, we note that $U^A_B(x,x) = \delta^A_B$, which completes
the proof.
\qed

Because the matrix (\ref{matrix}) satisfies (\ref{rep}),
so does its determinant. However, this does not give us
any new representation, because $\det U$ equals the Jacobian
(\ref{jac}). To see this, we note that (\ref{matrix}) can
be written as
\be
U^A_B = (\dd {x^B} + k_B) y^C \; ( \delta^A_C + c M^A_C ),
\ee
where
\be
M^A_B = y^A (\dd{y^B} + k_B) x^C \ddd h {x^C}.
\ee
Now,
\be
\log\ \det (1+cM) = \tr\ \log (1+cM)
= \sum_{n=1}^\infty {{{(-c)}^n} \over n} \tr\ M^n = 0,
\ee
because
\be
\tr\ M^n \propto y^A \dd{y^A} = 0.
\ee
We thus conclude that $\det U$ is independent of $c$.
When $c=0$,
\be
U^A_B = \pmatrix{
1 &0 \cr
\cr
(\ldots) & \displaystyle{\ddd\ymu\xnu} }
\ee
has a block triangular form, and its determinant is hence
\be
\det U = \det \ddd y x = \Jac y x,
\ee
as claimed.

{\em 3.}
Eq. (\ref{conf}) is the infinitesimal form of
a conformal field. Consider the infinitesimal diffeomorphism
\be
y^A = x^A + f^A.
\ee
To first order in $f$,
\be
\d_B y^A \approx \delta^A_B - k_B x^A + \d_B f^A,
\ee
\be
h(y,x) \approx \tr\ \log \Big( 1 + \ddd f x \Big)
\approx \tr \Big( \ddd f x \Big) = \d_\si f^\si,
\ee
and
\be
U^A_B(y,x) \approx \delta^A_B + (\d_B + k_B) f^A
+ c x^A \d_B \d_\si f^\si.
\ee
Thus,
\be
\phip^A(y) \approx \phi^A(x) - L_f \phi^A(x)
+ f^\mu \d_\mu \phi^A(x)
\approx U^A_B(y,x) \phi^A(x)
\ee
from which we deduce that
\be
L_f \phi^A(x) =  f^\mu \d_\mu \phi^A(x)
- \Big[ (\d_B + k_B) f^A + c x^A \d_B \d_\si f^\si \Big] \phi^B(x).
\ee
This is recognized as (\ref{conf}) in the vector case, with
\be
T^A_B \phi^C = - \delta^C_B \phi^A.
\ee

{\em 4.}
As any representation of $Diff(N)$, a conformal field yields
a representation of every subgroup by restriction.
The group of projective transformations (\ref{proj}) consists
of the diffeomorphisms of the form
\def\DelInv{\Delta^{-1}}
\be
y^A = \DelInv a^A_B x^B,
\qquad
\Delta = k_A a^A_B x^B,
\ee
where we have put $t=1$.
To calculate the form of the matrix (\ref{matrix}), we
need
\be
\d_B y^A = \DelInv (a^A_B - \DelInv a^A_C x^C k_D a^D_B ).
\ee
In particular,
\be
\d_\nu y^\mu = \DelInv a^\mu_\si (\delta^\si_\nu - M^\si_\nu),
\ee
where
\be
M^\mu_\nu =  \DelInv a^0_\nu (\xmu + (a^{-1})^\mu_\tau a^\tau_0 ).
\ee
($a^{-1}$ is the inverse of $(a^\mu_\nu)$.)
We note that
\be
\d_\nu \Delta = a^0_\nu
\ee
and
\be
\d_\nu M^\mu_\si = \DelInv a^0_\si (\delta^\mu_\nu - M^\mu_\nu).
\ee
Hence,
\bea
h(y,x) & = & \log \det \Jac y x \lf
& = & - N \log \Delta + \log \det a + \tr\ \log (1-M),
\eea
and
\be
\d_\nu h(y,x) = - N \DelInv a^0_\nu + \tr\ \d_\nu \log (1-M)
= -(N+1) \DelInv a^0_\nu.
\ee
Collecting all terms, we find that
\be
U^A_B(y,x) = \DelInv \Big\{ a^A_B +
(1+c(N+1)) \DelInv a^A_D x^D (k_B - \DelInv k_C a^C_B) \Big\}.
\label{sl}
\ee

Actually, (\ref{sl}) is the product of two factors, each of which
satisfies (\ref{rep}). In an obvious matrix notation, two
consequtive projective transformations take the form
\be
y = {{a x} \over {k^T a x}},
\qquad \qquad
z = {{b y} \over {k^T b y}}.
\label{prj}
\ee
The proof that
\be
\DelInv(z,y) \DelInv(y,x) = \DelInv(z,x),
\ee
reads
\be
(k^T b y) (k^T a x) = k^T {{a x} \over {k^T a x}} (k^T a x)
= k^T b a x.
\ee
That the other factor in (\ref{sl}) also satisfies
(\ref{rep}) can be verified in a similar fashion.
In particular, we have the following infinite-dimensional
$SL(N+1)$ representation for $c = -1/(N+1)$.
\be
\phip^A(y) = a^A_B \phi^B(x),
\ee
where $y$ is related to $x$ by (\ref{prj}).
We believe that the corresponding representations for
$c \ne -1/(N+1)$ are new.

{\em 5.}
In this letter we have only dealt with the local properties
of the diffeomorphism group. The name $Diff(N)$
then makes sense, because any manifold is locally diffeomorphic
to a subset of $\IR^N$. It would be interesting to determine
for which $N$-dimensional manifolds X
the group action be extended to $Diff\ X$, the
diffeomorphism group on $X$.
It is clear that the construction goes through for the group
$Diff_{Loc}\ X$ consisting of maps
with local support, reducing to the identity
map outside some fixed region.  Evidently,
\be
Diff_{Loc}\ X \subset Diff\ X,
\label{diff}
\ee
so we may consider the $Diff\ X$ representation induced from
a $Diff_{Loc}\ X$ conformal field. However, it is not clear to us
if any subtleties arise since both groups in (\ref{diff}) are
infinite dimensional. Indeed, (\ref{matrix}) is not
well defined on the torus, because it explicitly involves the
coordinates and not just their derivatives.

In \cite{Lar92b} we constructed first-order differential
operators (exterior derivatives), which commutes with the
action of $Vect(N)$. Preliminary calculations show that the
corresponding result holds on the group level.

A point worth noting is that a conformal field depends on two
parameters: $c$, which appears explicitly in (\ref{matrix}),
and $t$ which enters through (\ref{notation}). We may thus
denote it by $\phi(x,t;c)$. As this notation suggests, the
parameter $t$ in many ways behave as an extra dimension
(``time''), in addition to the $N$ (``space'') dimensions
labelled by $x$.
Moreover, conformal fields are useful
in the construction of $Vect(N)$ lowest-weight modules, which
may be relevant to quantum gravity \cite{Lar92c}.

We conclude by observing that
invariantly defined field-like objects (\ref{field}) are
scarce; previously, only tensor fields were known (spinor
fields are more complicated, because their
definition requires additional structure in the form of a
metric or a vielbein).
Presumably, no further field-like objects
exist, because $SL(N+1)$ is the largest finite-dimensional
subgroup from which a $Diff(N)$ module can be induced.
Therefore, we believe that conformal
fields have an important role to play in physics and geometry.

\vfill \eject


\begin{thebibliography}{99}

\bibitem{Rud74}
A. A. Rudakov, Math. USSR Izvestija 8 (1974) 836.

\bibitem{VGG75}
A. M. Vershik, I. M. Gelfand and M. I. Graev,
Russian Math. Surveys 30:6 (1975) 1.

\bibitem{Kir76}
A. A. Kirillov, Russian Math. Surveys 31:4 (1976) 57.

\bibitem{BeLe81}
J. N. Bernstein and D. A. Leites, Sel. Math. Sov. 1 (1981) 143.

\bibitem{Fuks87}
D. B. Fuks, Cohomology of infinite-dimensional Lie algebras
(New York and London: Plenum Press 1987).

\bibitem{Lar92a}
T. A. Larsson, to appear in Int. J. Mod. Phys. A (1992)
(hep-th/9207029).

\bibitem{Lar92b}
T. A. Larsson, submitted to Int. J. Mod. Phys. A (1992)
(hep-th/9207030).

\bibitem{Lar92c}
T. A. Larsson, submitted to Int. J. Mod. Phys. A (1992).

\bibitem{Kir76}
A. A. Kirillov, Elements of the theory of representations
(Springer 1976).

\end{thebibliography}
\end{document}